\documentclass[useAMS,usenatbib]{mn2e}

\usepackage[latin1]{inputenc}
\usepackage[pdftex]{graphicx}
\usepackage{epstopdf}
\usepackage{graphicx}\usepackage{subfigure}
\usepackage{rotate}
\usepackage{color}

\newcommand       \degree      {\rm ^o}

\title[Characterization of the optical and X-ray properties of the northwestern wisps in the Crab nebula]{Characterization of the Optical and X-ray Properties of the Northwestern Wisps in the Crab Nebula}
\author[T.~Schweizer {\it et al.}]
{T. Schweizer$^{1}$\thanks{E-mail:tschweiz@googlemail.com},
N. Bucciantini$^{2,3}$, W. Idec$^{1,4}$, K. Nilsson$^{5}$, A. Tennant$^{6}$,
\newauthor M.C. Weisskopf$^{6}$ and R. Zanin$^{7}$\\
$^{1}$Max-Planck-Institute for Physics, Foehringer Ring 6, 80805 Munich, Germany\\
$^{2}$INAF - Osservatorio Astrofisico di Arcetri, L.go E.~Fermi 5, 50125, Firenze, Italy\\
$^{3}$INFN - Sezione di Firenze, Via G.~Sansone 1, 50019 Sesto Fiorentino, Firenze, Italy\\
$^{4}$Department of Astrophysics, University of \L \'od\'z, ul. Pomorska 149/153, 90-236 \L \'od\'z, Poland\\
$^{5}$Finnish Centre for Astronomy with ESO (FINCA), University of Turku, 20, FI-21500 , Finland\\
$^{6}$Space Science Department, NASA Marshall Space Flight Center, ZP12, Huntsville, AL 35812\\
$^{7}$Universitat de Barcelona, Departament d'Astronomia i Meteorologia, E-08028 Barcelona, Spain}

\begin{document}

\date{}

\pagerange{\pageref{firstpage}--\pageref{lastpage}} \pubyear{2012}

\maketitle

\label{firstpage}

\begin{abstract}
We have studied the wisps to the north-west of the Crab pulsar as part
of a multiwavelength campaign in the visible and in X-rays.
Optical observations were obtained using the Nordic Optical Telescope
in La Palma and X-ray observations were made with the Chandra X-ray Observatory.
The observing campaign took place from 2010 October until 2012 September.
About once per year we observe wisps forming and peeling off from (or
near) the region commonly associated with the termination shock of the pulsar wind.
We find that the exact locations of the northwestern wisps in the optical and in
X-rays are similar but not coincident, with X-ray wisps preferentially located closer to the pulsar.
This suggests that the optical and X-ray wisps are not produced by the same particle distribution.
Our measurements and their implications are interpreted in terms of a
Doppler-boosted ring model that has its origin in magnetohydrodynamic (MHD) modelling.
While the Doppler boosting factors inferred from the X-ray wisps are
consistent with current MHD simulations of pulsar wind nebulae (PWN), the optical boosting
factors are not, and typically exceed values from MHD simulations by about a factor of 3.
\end{abstract}

\begin{keywords}
ISM: radiation mechanisms: non-thermal - pulsars:individual: Crab -
ISM: supernova remnants
\end{keywords}

\section{Introduction}

The Crab nebula is one of the most studied targets in the sky as it is
bright and observable over a very broad spectral range.
The nebula is a remnant from a supernova explosion that was observed on earth in 1054 CE.
Located at a distance of $\sim$2~kpc, the system is powered by a
pulsar of spin-down luminosity $L \sim 5\times10^{38}$ erg s$^{ -1}$ and period $P \sim 34$~ms.
The history and general properties of the system are nicely summarized in the review by \citet{Hester08}.

Amongst the most prominent features of the inner nebula are a jet to the south and a counter-jet to the north, a torus and a rich variable ``wisp''
structure \citep{Scargle1969} more prominent to the north-west (NW) than the south-east.
In the past decade, there have been several detailed observations in
various wavelengths that observed the jets, found a knot at the
beginning of the southern jet close to the pulsar (about $0.7\:  arcsec$), and
reconfirmed the wisps \citep{Hester95, Hester02, Hester08}.
Observations with the Hubble Space telescope (HST) had shown that the
optical wisps form and dissipate over time-scales of months \citep{Hester02}.
In addition, variable (in both position and time) X-ray-emitting knots
also are present, especially to the southeast of the pulsar \citep{Weisskopf2012}.
A comparison between optical and radio wisps may be found in \citet{Bietenholz04}.

The discovery of $\gamma$-ray flaring in 2010 September \citet{Tavani2011,
Abdo2011} stimulated a renewed interest in the Crab nebula.
Here we present the results of two observing campaigns, one in the
visible and one in X-rays inspired by the search for the origin of the $\gamma$-ray flaring.
In this paper we concentrate on characterizing the wisps to the NW of the pulsar.
As part of the campaign, we also obtained data during during the
$\gamma$-ray flaring activity of 2011 April \citep{Buehler2012}.
However, we see no direct correlation between the appearance of either
an optical or an X-ray wisp and the $\gamma$-ray flaring.

We describe the optical (\S\ref{s:optobs}) and X-ray (\S\ref{s:xobs}) observations.
Data reduction and derived physical parameters are presented as
follows: wisp radial profiles (\S\ref{s:wrp}), wisp velocities
(\S\ref{s:wv}) and wisp azimuthal profiles (\S\ref{s:wap}).
In \S\ref{s:mft} we make use of a Doppler-boosted ring model to describe the observations and discuss implications.
We briefly summarize our findings in \S\ref{s:sr}.

\section{The Observations}

\subsection{The optical observations} \label{s:optobs}

\begin{figure}
\centering
\includegraphics[width=0.5\textwidth, keepaspectratio]
	{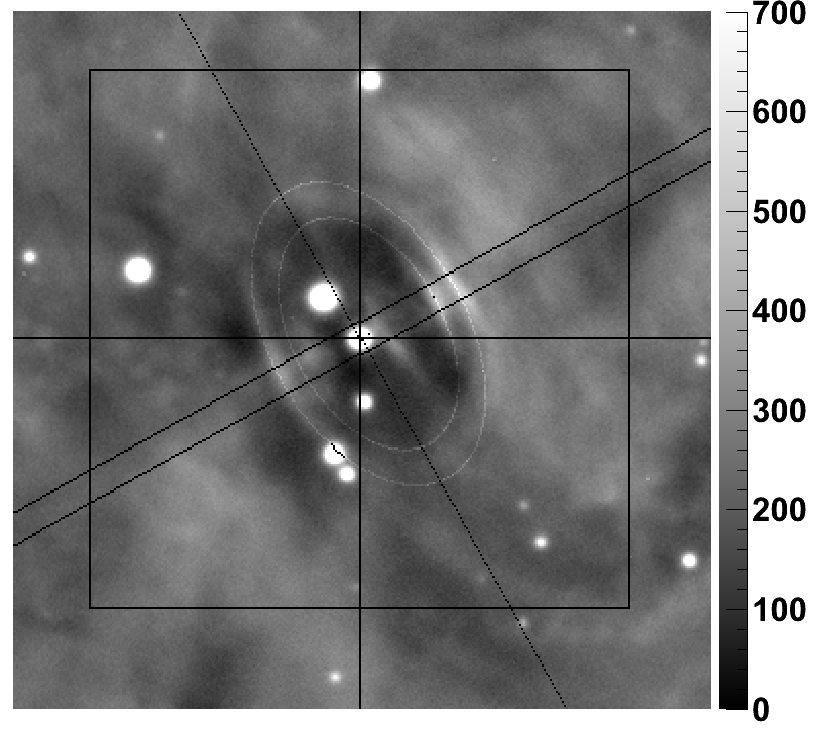}
\caption{\label{fig:Crab_NOT} Optical image of the inner portion of
the Crab nebula taken on 2011 April 13  and during 
a $\gamma$-ray flare \citep{Buehler2012}.
North is up and east is to the left.
The square covers a field of view of $50\:  arcsec \times 50\:  arcsec$
and is centred on the pulsar. The two parallel lines at a position
angle of $-61 ^\circ$ {measured east from north} outline the region used to measure radial profiles.
Two ellipses are shown that pass through the most prominent wisps
at $8.0\:  arcsec$ and $10.1\:  arcsec$ NW of the pulsar.
The aspect ratio of the ellipses (for all images) is 0.6 and their centre is offset
from the pulsar by $0.8\:  arcsec$ along a line (to the
NW) that would  connect the pulsar to the wisp peak emission.
For display purposes we have removed low frequency variations in the image.
The intensity units are arbitrary.
}
\end{figure}

\begin{table}
\begin{center}
\caption{Optical observation dates and seeing resolution
\label{tab:optical}}
\begin{tabular}{llc}
\hline \hline
MJD$^a$ & Date$^b$ & Seeing$^c$  \\
\hline
55518.1   &   2010/11/18   & 0.49 \\
55532.0   &   2010/12/02   & 0.69 \\
55557.9   &   2010/12/27   & 0.83 \\
55564.1   &   2011/01/03   & 0.87 \\
55570.0   &   2011/01/09   & 0.59 \\
55598.8   &   2011/02/06   & 0.62 \\
55612.9   &   2011/02/20   & 0.70 \\
55647.9   &   2011/03/27   & 0.49 \\
55664.9   &   2011/04/13   & 0.78 \\
55810.2   &   2011/09/06   & 0.54 \\
55863.2   &   2011/10/29   & 0.99 \\
55881.1   &   2011/11/16   & 0.60 \\
55922.1   &   2011/12/27   & 0.61 \\
55954.9   &   2012/01/28   & 1.05 \\
55986.9   &   2012/02/29   & 0.72 \\
55998.9   &   2012/03/12   & 0.96 \\
56013.9   &   2012/03/27   & 0.60 \\
56025.9   &   2012/04/08   & 0.75 \\
56033.9   &   2012/04/16   & 1.02 \\
56140.2   &   2012/08/01   & 0.82 \\
56141.2   &   2012/08/02   & 0.82 \\
56153.2   &   2012/08/14   & 0.52 \\
56159.2   &   2012/08/20   & 1.10 \\
56176.2   &   2012/09/06   & 0.75 \\
\hline
\end{tabular}
\end{center}
$^a$ Modified Julian Date of start of observation.\\
$^b$ UTC date.\\
$^c$ Seeing FWHM in arcsec.\\
\end{table}

Optical data were obtained using the $2.56$m Nordic Optical Telescope
(NOT) located in the Observatorio del Roque de los Muchachos on the Canary island La Palma.
Altogether 24 images were taken from 2010 November until 2012 September
using the Andalucia Faint Object Spectrograph and Camera (ALFOSC).
The instrument employs a $2048 \times 2048$ E2V chip with a gain of
$0.327e^-$/ADU, a  readout noise of $4.2 e^-$ and a scale of
$0.19 \:  arcsec \: pixel^{-1}$ giving a field of view of $6.5 \times 6.5$ arcmin.
Two to four exposures of $200 - 300$s were made through the I-band
filter (NOT filter \#12, $\lambda_{c}$ around $800$nm) at each epoch.
The effective seeing resolution depends on observing conditions
and ranges between $0.49 \: arcsec$ and $1.1\:  arcsec$.
Table~\ref{tab:optical} lists the dates and seeing resolution for each
optical observation.
Fig.~\ref{fig:Crab_NOT} shows one of the 24 optical images.

The images were processed first by subtracting the average bias image and then
dividing by a twilight-sky flat-field image.
The fringe pattern was removed using a fringe correction image created
from archival NOT data taken using the same instrument configuration.
After scaling and subtracting the fringe correction image, the fringe pattern was no longer visible.
Individual frames were then registered using 8 stars in the vicinity
of the pulsar and then averaged.

We checked the NOT pixel scale
using a HST-legacy archive image of the Crab pulsar obtained with the Wide Field
and Planetary Camera 2 (WFPC2) on 1995, August 14. Using the Planetary
Camera  portion of the image we selected two pairs
of unsaturated stars within 25 arcsec of the pulsar and determined
the distance between the two stars in each pair using the $0.05\:  arcsec\: 
pixel^{-1}$ pixel scale of the
image. Measuring the distances of the same pairs in the NOT image we
arrived at NOT pixel scale
 of $0.1902$ and $0.1904\:  arcsec\:  pixel^{-1}$. Hence the adopted pixel scale
 $0.19\: \:  arcsec\:  pixel^{-1}$ is expected to be accurate to
within 0.2\%.

Sky background was subtracted from the average frames by fitting two
models to the ``pure sky'' regions around Crab: a constant and a low-order polynomial.
The two methods generally agreed to better than $1.2$\% when
extrapolated to the
location of the pulsar.
We use the constant in what follows.
Flux calibration used the comparison star sequence given in \citet{Sandberg2009}.
We used star 3 with I-band magnitude I = $16.26 \pm 0.02$ from their calibration
sequence to determine the zero point, $ZP$, by fitting a point spread
function (PSF) model created from star 1 of \citet{Sandberg2009} to star 3.
The conversion from sky-subtracted counts, $N$ (ADU), to physical
fluxes, $F$ ($\mu Jy$ per pixel), is then
\begin{equation}
F = 2550 * 10^{-0.4(ZP - 2.5 \log N)}\ .
\end{equation}
Since the calibration was obtained from a star in the Crab image,
variations in atmospheric transparency are effectively accounted for
and removed.

To check the accuracy of the calibration and background subtraction we
studied the surface brightness of a relatively calm region close to the
pulsar over our monitoring period. Using the 24 flux calibrated images
we computed an average image and a sigma image, where in the latter each
pixel value represents the rms flux variability at that pixel. Examining
this image we identified a $2.8\:  arcsec \times 2.8\:   arcsec$ region $37\:   arcsec$ South
of the pulsar with a surface  brightness close to that of the wisps
(140 $\mu$Jy per $arcsec^{2}$) but with a small rms scatter compared
to other regions near the pulsar. The rms scatter of the surface
brightness from this region was 3.3\%, somewhat larger than we would
have expected due to inaccuracies in calibration and sky subtraction.
Thus, some time variability of the Crab nebula may still be present,
yet not at a level significant enough to alter our findings.

\subsection{The X-ray Observations} \label{s:xobs}

\begin{figure}
\centering
\includegraphics[width=0.5\textwidth, keepaspectratio]
	{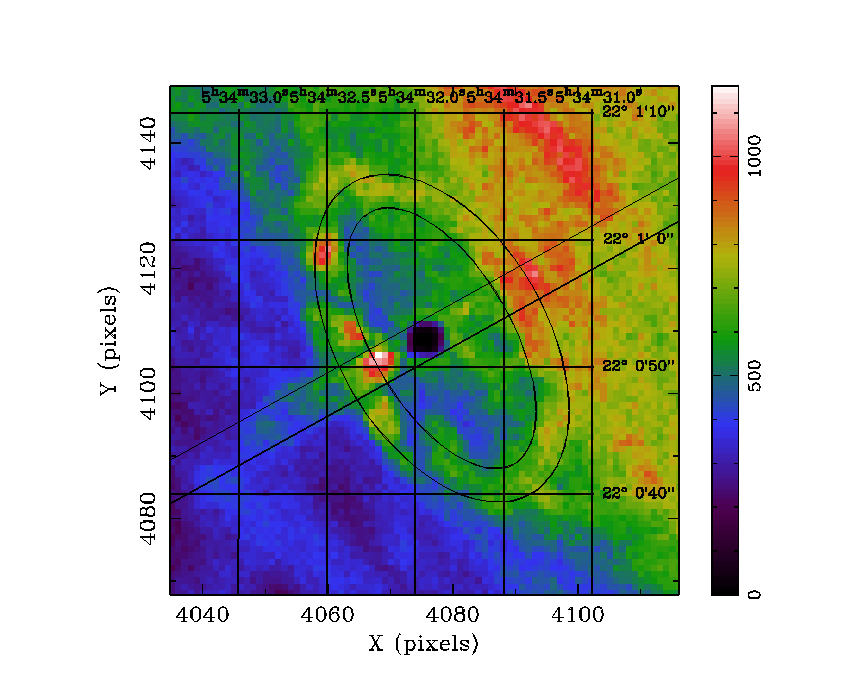}
\caption{\label{fig:Crab_Chandra}This Chandra X-ray image of the inner
  portion of the Crab nebula is the sum of the observations from the
  2011 April, 14 to 28 during the $\gamma$-ray flare. The
  axes show RA and Dec (J2000) and ACIS detector pixels ($0.492\:  arcsec
  \: pixel^{-1}$).
The colour scale is in counts per ACIS pixel in the 0.5-8 keV band.
The black feature with zero counts is the burned out image of the pulsar.
The two parallel lines show the region from which the X-ray radial profiles
were extracted and are the same as those in Fig.~\ref{fig:Crab_NOT}.
The two ellipses show the region from which the azimuthal X-ray
profile  was extracted.
}
\end{figure}

Subsequent to the discovery of $\gamma$-ray flaring in 2010 September
\citep{Tavani2011, Abdo2011} a series of Chandra X-ray Observatory
observations was initiated on approximately a monthly cadence.
Five Target of Opportunity observations were performed during the
2011 April $\gamma$-ray flare and the average of those 5 images is shown in Fig.~\ref{fig:Crab_Chandra}.
Chandra observations have continued, now as part of the Chandra
general observer programme, yielding a sequence which covers approximately two years.
As with our optical observations, the coverage is not continuous.
Since Chandra is not allowed to point within $45\degree$ of the Sun, there is
a 90 day interval during the summer when the Crab may not be observed.
A list of the Chandra observations used for this paper is given in Table~\ref{tab:xrayobs}.

All data were taken with the back-illuminated Advanced CCD Imaging
Spectrometer S3
(ACIS-S3)\footnote{http://asc.harvard.edu/proposer/POG/html/chap6.html}
detector at the focus of the telescope in the energy band 0.5-8 keV.
Owing to the Crab's high flux, the observations were taken with the
shortest possible frame time of $0.2$~s to minimize pileup.
Even with this frame time, pileup impacts the brighter regions of the nebula.
Telemetry saturation is also a issue with such a bright source, so we
restricted data to a $300 \times 300$~ACIS-pixel ($\approx
150\:  arcsec\times 150\:  arcsec$) region that covers the bright part of the nebula.
The integration time was five ks for monitoring observations and 10 ks
for the 5 observations following the 2011 April $\gamma$-ray flare.
Due to telemetry saturation the effective exposure time was $600$ or $1200$~s (Table~\ref{tab:xrayobs}).

For the first three observations (Table~\ref{tab:xrayobs}) the dither
amplitude of the spacecraft was set to zero.
Since pileup characteristics may be different if the pulsar image is
at the middle, edge or corner of a pixel, we then used a small ($1\:  arcsec$ amplitude) dither for the remaining observations.
This dither ensures that the pulsar and other sharp features in the nebula are averaged over a few pixels.

Level 2 event files were created using the CIAO tool {\tt
acis\_process\_events} using the EDSER option \citep{Li2004} to improve the subpixel positioning.
Due to severe pileup the pulsar does not appear in the image (Fig.~\ref{fig:Crab_Chandra}).
We did not simply rely on the Chandra aspect solution, but
re-registered images making use of the readout trail  (out-of-time image of the pulsar)  to constrain one
dimension and the burned out image of the pulsar to constrain the other coordinate.

Due to the different roll orientations of the observatory, the readout
trail could introduce some bias in
the azimuthal distribution. We removed this trail by computing the
average number of counts per pixel due to the trail.
Then, based on the actual number of counts seen in a pixel and our estimate of
the number due to the readout streak, we randomly rejected events.
Since this method is statistical, it is not perfect, but it will remove
the principle brightness peak in the data due to the trail itself.

\begin{table}
\begin{center}
{\caption {Time-ordered list of Chandra observations with exposure time used for this paper. \label{tab:xrayobs}}}
\begin{tabular}{lllr} \\ \hline
ObsID$^a$ & MJD$^b$ & Date$^c$ & Time$^d$ \\ \hline
13139 & 55467.2 & 2010/09/28 & 600 \\
13146 & 55497.7 & 2010/10/28 & 600 \\
13147 & 55528.5 & 2010/11/28 & 600 \\
13204 & 55576.0 & 2011/01/15 & 600 \\
13205 & 55608.7 & 2011/02/16 & 600 \\
13206 & 55635.2 & 2011/03/15 & 600 \\
13207 & 55663.6 & 2011/04/12 & 600 \\
13150 & 55665.0 & 2011/04/14 & 1200 \\
13151 & 55665.6 & 2011/04/14 & 1200 \\
13152 & 55667.4 & 2011/04/16 & 1200 \\
13153 & 55673.0 & 2011/04/22 & 1200 \\
13154 & 55679.3 & 2011/04/28 & 1200 \\
13208 & 55788.4 & 2011/08/15 & 600 \\
13209 & 55819.2 & 2011/09/15 & 600 \\
13210 & 55849.2 & 2011/10/15 & 600 \\
13750 & 55892.7 & 2011/11/27 & 600 \\
13751 & 55938.1 & 2012/01/12 & 600 \\
13752 & 55967.4 & 2012/02/10 & 600 \\
14416 & 56005.3 & 2012/03/19 & 600 \\
13754 & 56039.1 & 2012/04/22 & 600 \\
13755 & 56155.6 & 2012/08/16 & 600 \\
13756 & 56181.9 & 2012/09/11 & 600 \\
\hline
\end{tabular}
\end{center}
$^a$ Chandra observation identifier.\\
$^b$ Start of observation MJD.\\
$^c$ UTC date.\\
$^d$ Approximate effective exposure time in seconds. \\
\end{table}

\begin{figure}
\centering
\includegraphics[width=0.5\textwidth, keepaspectratio]
        {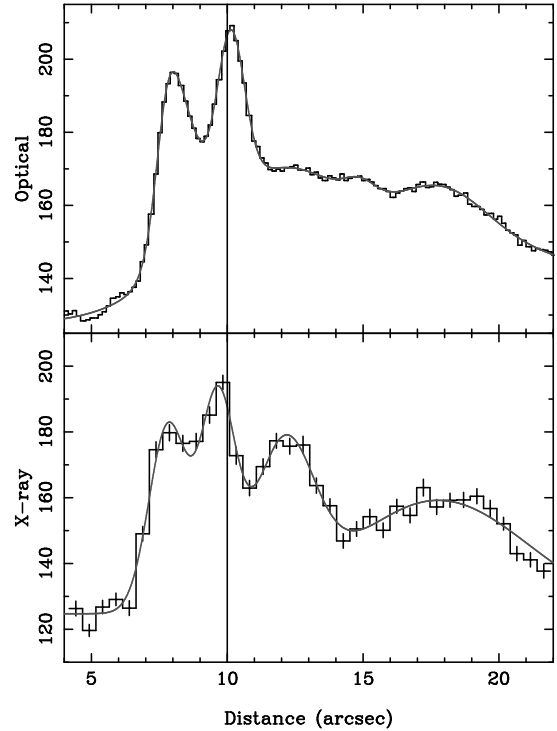}
\caption{The upper panel shows the
radial brightness distribution for the optical data for 2011 April 13 ($\mu Jy/arcsec^{2}$).
The bottom panel shows the corresponding X-ray radial distribution for 
the average from 2011 April 14 until April 28 (counts per ACIS pixel).
The solid lines denote a model fit based on a number of Gaussians.
The vertical line at $10\:  arcsec$ is drawn to emphasize that the optical and X-ray peaks do not exactly coincide.
\label{fig:April_radial_profiles}
}
\end{figure}

\begin{figure}
\centering
\includegraphics[width=0.5\textwidth, keepaspectratio]{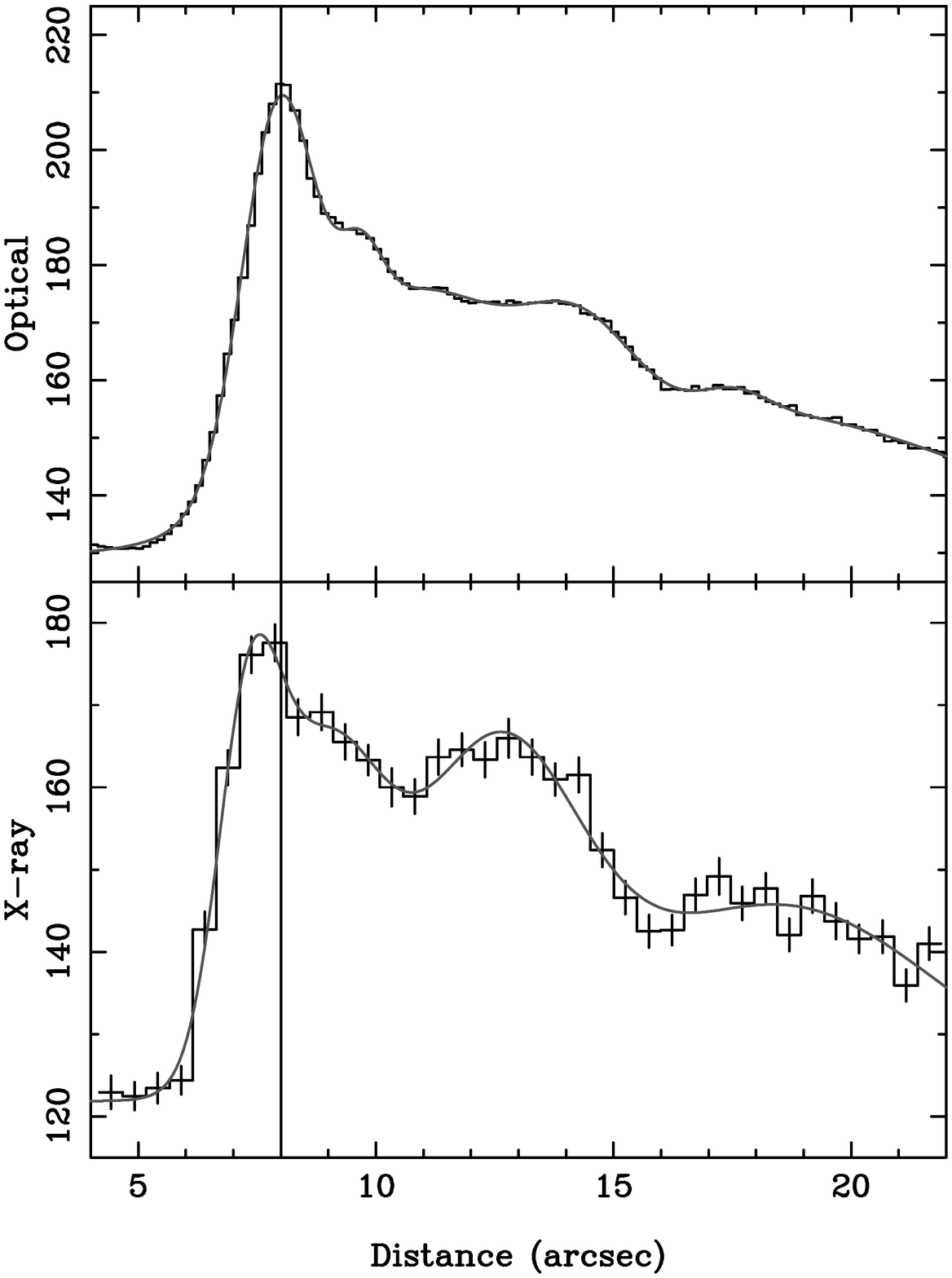}
\caption{The upper panel shows the radial profile for the average of
  all the optical observations (in units of $\mu Jy/arcsec^{2}$). The
  lower panel shows  the average radial profile for all the X-ray data
 (photon counts). The X-ray average does not include
  Obs\-IDs 13150 to 13154 to avoid biasing the
 average because of the concentrated coverage in 2011 April and to
maintain a similar sampling cadence as for the optical data.
The vertical line at $8\:  arcsec$ shows that the brightest and closest
optical and X-ray peaks do not exactly coincide.
}
\label{fig:average_radial_projections}
\end{figure}

\section{Data analysis} \label{s:da}

Images were analysed to characterize the shape of the wisps in both
the radial direction from the pulsar and angular extent about the pulsar.
Our analysis to determine the radial distribution through the wisps
restricted data to the narrow strip  $3\:  arcsec$ wide with position angle $-61\degree$ 
and is shown in Figs.~\ref{fig:Crab_NOT} and \ref{fig:Crab_Chandra}.
Optical and X-ray data were binned in radial bins of $0.11\:  arcsec$ and
$0.492\:  arcsec$ respectively. The centre of the ellipse is assumed to
be the same for all wisps. Significant differences
($<0.1\:  arcsec$) cannot be
detected within the resolution of individual optical or X-ray images.
For azimuthal distributions we utilized $0.5\:  arcsec$-wide
elliptical annuli, on the sky and centred on the peak of the radial
distribution. 

\begin{figure}
\centering
\includegraphics[width=0.5\textwidth,keepaspectratio]
	{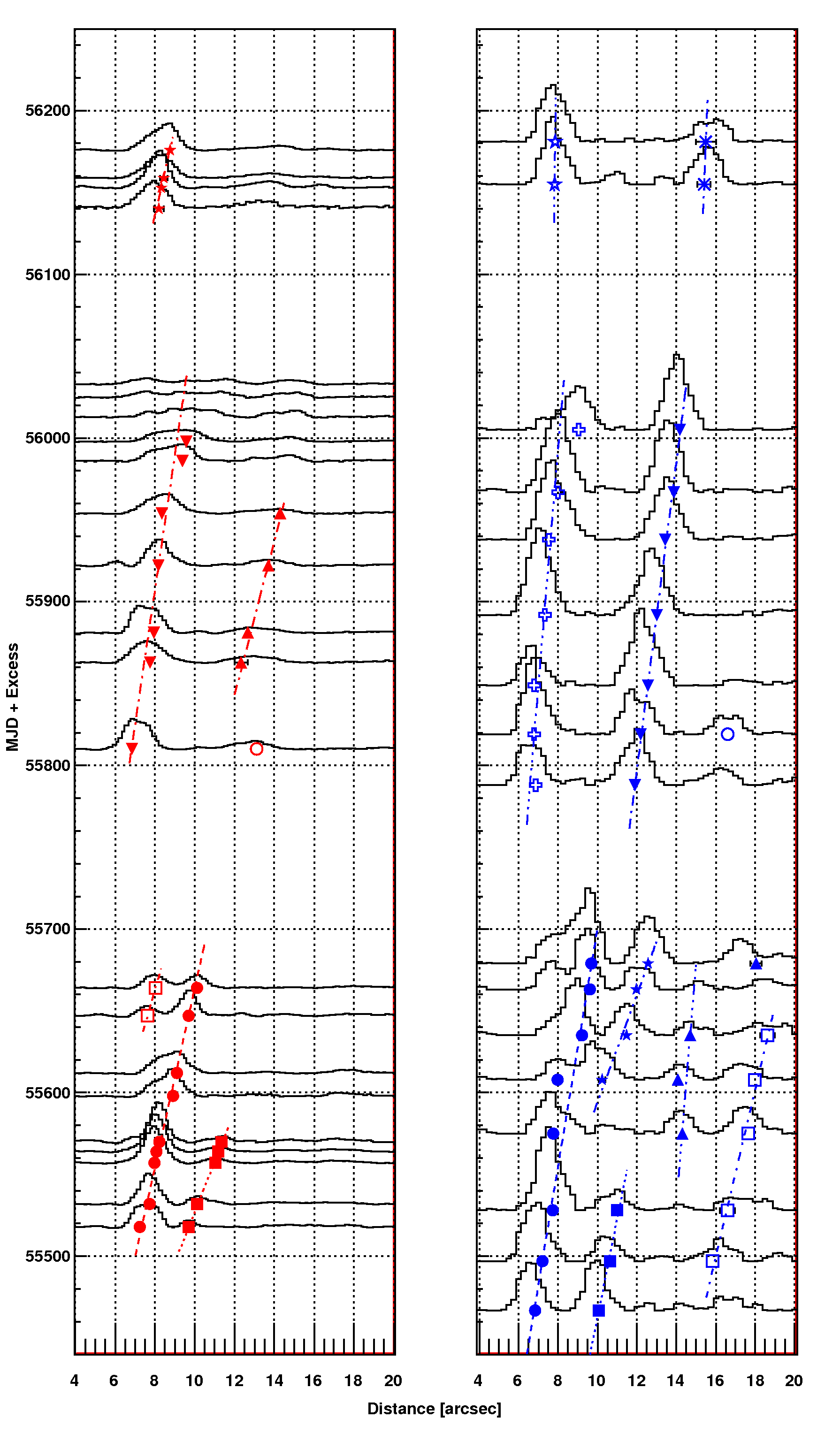}
\caption{23 optical and 17 X-ray radial profiles with
  low-frequency terms removed are plotted as a function of the distance NW of the pulsar.
For clarity when one or more observations occur very close in time, we omit them from the figure.
The omitted observations are optical - 56140.2  and X-ray -  55665.0,
55665.6, 55667.4, 55673.0 and 56039.1.  For display purposes, low
frequency  terms have been removed using an algorithm from the ROOT-analysis package
\citep{Background}.
The symbols are placed on the distance axis to show the position of the
peak of the Gaussian fitted to the original profile.
So doing gives the illusion that the symbols are offset from the peak value.
The lines, simply best guesses, are an interpretation of the time evolution of the position
of a particular peak.
}

\label{fig:Crab_optical_and_xray_profile_stack}
\end{figure}

\begin{figure}
\centering
\includegraphics[width=0.5\textwidth,keepaspectratio]
	{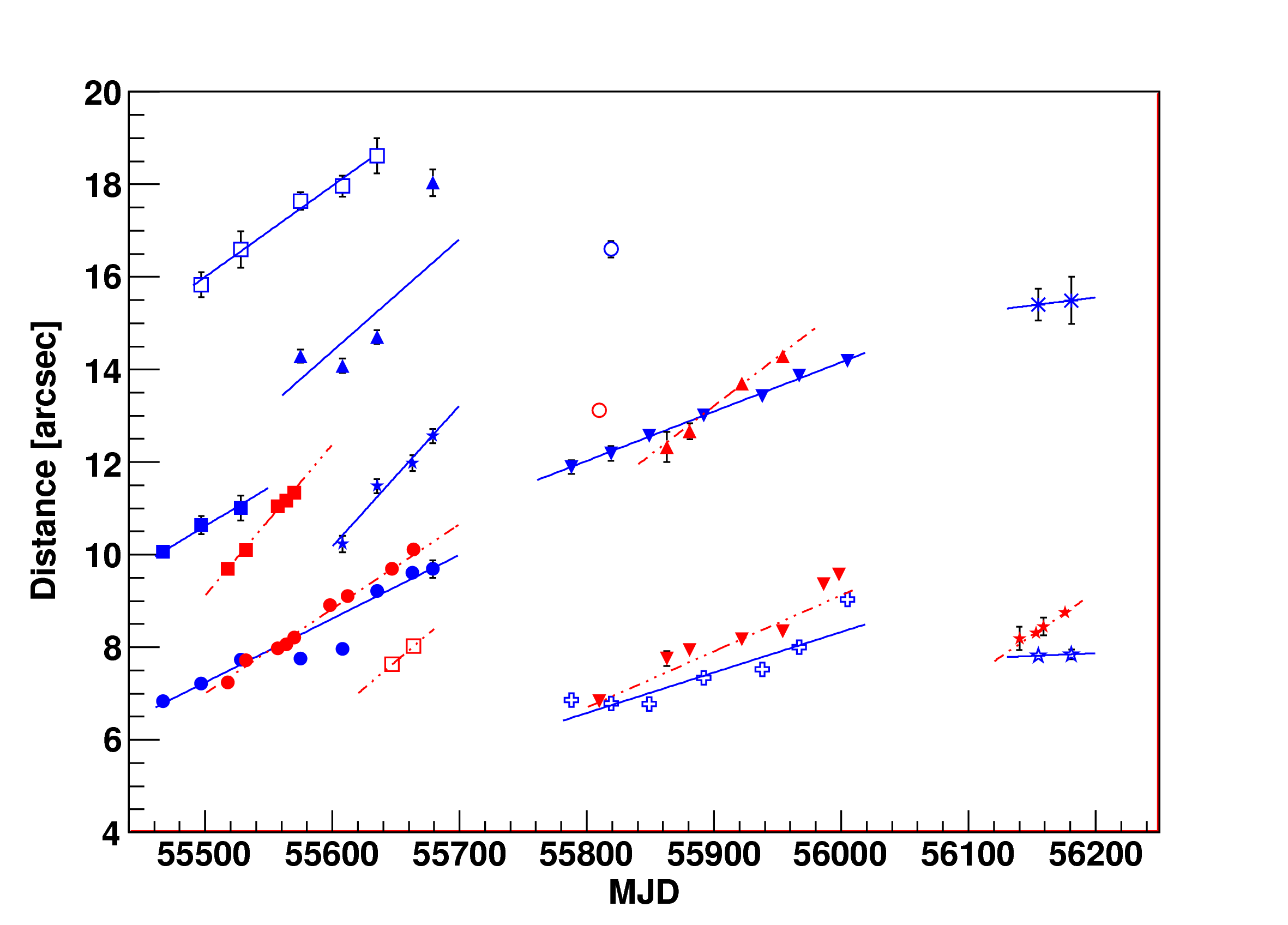}
\caption{The radial positions of the wisp peaks as a function of time.
Red and blue are used to indicate an optical or an X-ray wisp respectively.
The symbols are the same as in Fig.~\ref{fig:Crab_optical_and_xray_profile_stack}.
}
\label{fig:Crab_optical_and_xray_peaks_vs_time}
\end{figure}

\begin{table}
\begin{center}
{\caption {Velocities (in units of c) of the wisps identified in
    Figs.~\ref{fig:Crab_optical_and_xray_profile_stack}
and ~\ref{fig:Crab_optical_and_xray_peaks_vs_time}.
}
\label{tab:velocities}}
\begin{tabular}{lccc}
 symbol      & $ arcsec/:d^{-1}$ &(v/c)$^a$ & (v/c)$^b$ \\
\hline
 \multicolumn{4}{c}{Optical (red)} \\
 \\
 circle        & 0.018         & 0.21     & 0.29      \\
 box solid     & 0.033         & 0.38     & 0.44      \\
 triangle up   & 0.021         & 0.24     & 0.32      \\
 triangle down & 0.012         & 0.14     & 0.21      \\
 star          & 0.019         & 0.22     & 0.30     \\
 \\
 \multicolumn{4}{c}{X-rays (blue)} \\
 \\
 circle        & 0.014         & 0.16     & 0.24      \\
 box solid     & 0.017         & 0.19     & 0.28      \\
 box open      & 0.020         & 0.23     & 0.31      \\
 star solid    & 0.031         & 0.35     & 0.42      \\
 triangle up   & 0.024         & 0.28     & 0.36      \\
 triangle down & 0.011         & 0.12     & 0.19      \\
 cross         & 0.009         & 0.10     & 0.16      \\
\end{tabular}
\end{center}
$^a$ Apparent velocity on the sky. \\
$^b$ Deprojected velocity. \\
\end{table}

\subsection{Analysis of the radial profile of the wisps}  \label{s:wrp}

Throughout this paper we refer to ``radial profiles''. This profile is simply the intensity distribution, within the parallel lines in Figs.~\ref{fig:Crab_NOT} and \ref{fig:Crab_Chandra} measured outwards from the pulsar towards the NW.
We fit a constant plus a number of Gaussians to both optical and X-ray radial profiles.
Initially a Gaussian is located at each peak seen in the profile.
The model was then fitted using a least squares fitting method that allowed
all parameters to vary.  After the fit had converged, we then visually
inspected to ensure that the main peaks had been properly located and if 
not the model would be modified, either by adjusting parameter values 
or adding an additional Gaussian, and then refitted.

An example of an optical and an X-ray radial profile measured close in
time is  shown Fig.~\ref{fig:April_radial_profiles}.
We see two prominent peaks in the optical and the presence of at least
three  peaks in the X-ray profile.
In general, the peak closest to the pulsar is most usually
 identified with the location of the termination shock.
Fig.~\ref{fig:April_radial_profiles} demonstrates that the X-ray peak
at $10\:  arcsec$  is located at a slightly different (and smaller)
distance than the peak in the optical.
The exposures in optical and X-rays in the figure are not precisely simultaneous and the average time difference is about 8 d.
However, even if we account for the apparent radial motion (\S~\ref{s:wv}), the X-ray wisp is still closer to the pulsar than the nearest optical wisp.

Fig.~\ref{fig:average_radial_projections} top plots the radial profile of
the average of all the optical observations.
Notice the strong peak at $8\:   arcsec$ showing that the optical wisps are
brightest at this distance from the pulsar and then fade at larger distances.
Fig.~\ref{fig:average_radial_projections} bottom shows the same plot in X-rays averaged with the same sampling interval as for the optical.
The innermost X-ray arc, at $\sim 7.5~ arcsec$, is brightest, and closer to the pulsar than in the optical. Similarly, the further from the pulsar, the fainter the average intensity in the radial profile.

\subsection{The radial evolution of the wisps as a function of time } \label{s:wv}

Fig.~\ref{fig:Crab_optical_and_xray_profile_stack} is a graphical
representation showing most of the optical and X-ray radial projections
as a function of the angular distance from the pulsar.
For display purposes only, low frequency terms have been subtracted. The algorithm applied was the TSpectrum background subtraction algorithm from the ROOT-package \citep{Background}.

In Fig.~\ref{fig:Crab_optical_and_xray_profile_stack} we also trace the outward motion of a particular peak by
drawing a line through the position of what appears to be the same
peak but at different times. There appears to be possibly five distinct
progressions of peaks in the optical data and possibly seven in X-rays.
We have not attempted to connect progressions across summer gaps,
but one could do so. As the slopes of the lines indicating the outward
progression are different, so are the inferred velocities.
These are given in Table~\ref{tab:velocities} where we list the
measured quantity i.e. the
apparent velocity on the sky in
$ arcsec/day$, and, assuming a distance of 2 kpc, the deprojected physical velocity.
For deprojection we use an inclination
angle of 
 $57^{^\circ}$ (\citet{Weisskopf2012} and references therein).
Fig.~\ref{fig:Crab_optical_and_xray_peaks_vs_time} also compares the
motion of the peaks in the radial profiles as a function of time.
The figure indicates that most of the time an X-ray wisp appears
close to the pulsar, so does an optical wisp that is slightly further away from the pulsar.

\begin{figure}
\centering
\includegraphics[height=0.5\textwidth, keepaspectratio, angle=270]
	{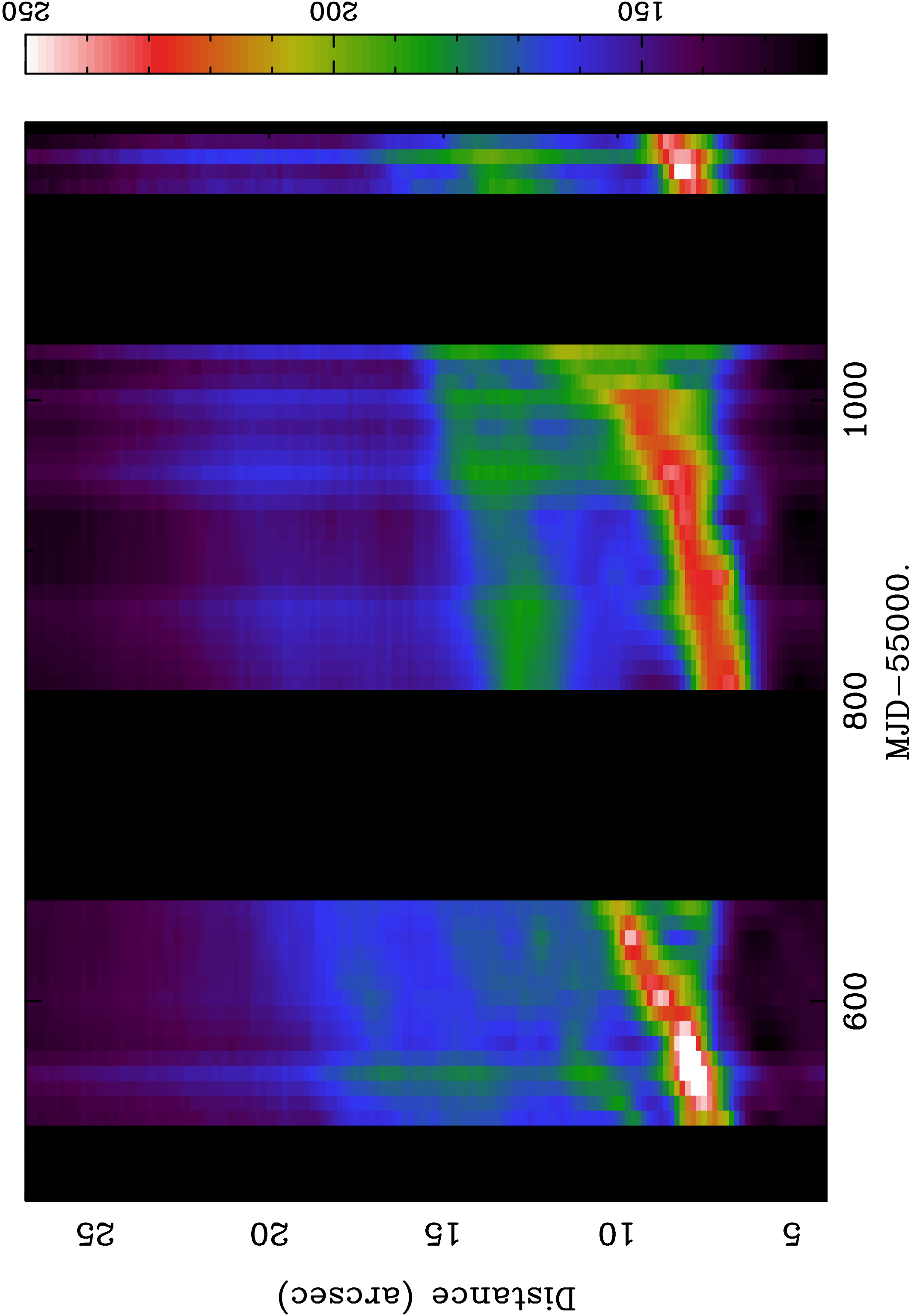}
\includegraphics[height=0.5\textwidth, keepaspectratio, angle=270]
	{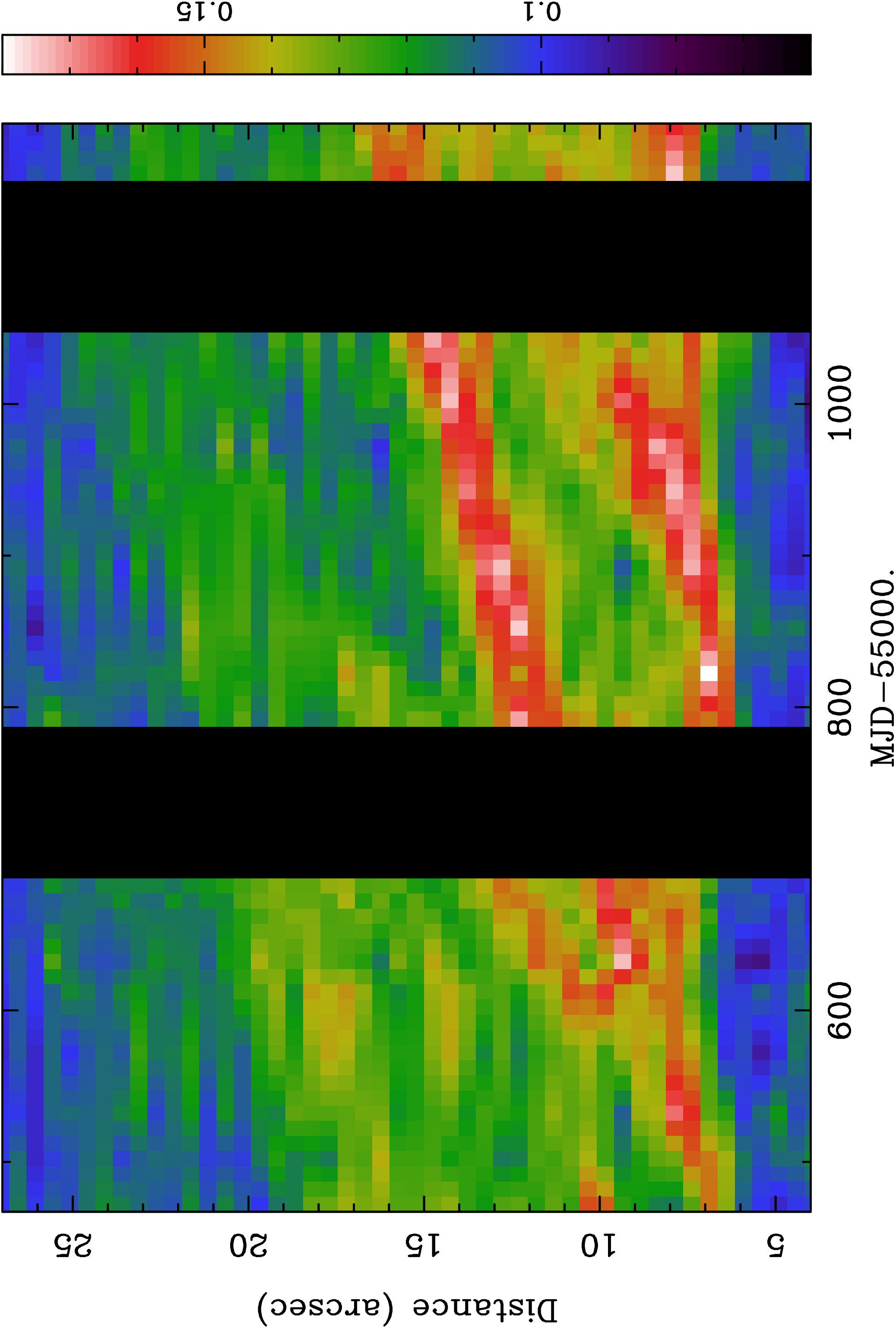}
\caption{The two panels compare the radial evolution of the optical
  (upper panel) and X-ray (lower panel) wisps. For clarity, data were
  interpolated between observations but not across the large gaps
  imposed by sun constraints.
  The unit for the colour scale is $\mu  J/arcsec^{2}$ for the upper
  panel  and counts for the lower panel.
}
\label{fig:Crab_wisp_evolution}
\end{figure}

Fig.~\ref{fig:Crab_wisp_evolution} presents a different yet equally
interesting, picture and comparison of the evolution of the peak
fluxes in the azimuthal distributions as a function of time.
To construct this figure we used a time spacing of 10 d.
If an observation took place at any time within the 10 d interval
the data were included as a column in the figure.
For short gaps in the time sequence, we performed a linear average of
the closest columns that contain data.
Thus, if there was a two column gap, we would add two thirds of the
previous observation to one third of the following observation to fill in the first missing column.
For the second missing column, the weights were reversed.
If the wisps in two data sets overlap, this method will cause the wisp to
appear to move from the location in the first data set to the location
seen in the second across the gap.
If the features do not overlap, then features in the first data set will appear
to fade as new features appear to grow.
We have determined that the gap in the summer break is too long for
our interpolation method to work and thus it is empty.

All the features seen in Fig.~\ref{fig:Crab_optical_and_xray_profile_stack} can
also be seen pictorially in Fig.~\ref{fig:Crab_wisp_evolution}.
Note that the width of neighbouring peaks in the X-ray and optical do not appear to be correlated.
In the X-ray portion of Fig.~\ref{fig:Crab_wisp_evolution} we also see
the slow outward motion of the outer boundary where the colour changes from green to blue.
This boundary is close to $20\:  arcsec$ from the pulsar for the earliest observations,
yet moves to near $25\:  arcsec$ by the end of the sequence.
This outward motion of the boundary roughly matches the apparent outward
motion of the brighter wisps.
Although the bright wisps can be seen in both the X-ray and the optical,some X-ray peaks do not appear to have a nearby optical companion. 
Furthermore,  the positions of the optical wisps
do not precisely align with the positions of the X-ray wisps.
That would seem to imply that the individual evolution of optical and X-ray wisps is different, but, becuse of their proximity they are related.
Fig.~\ref{fig:Crab_wisp_evolution} also makes it clear that new wisps form in the inner region roughly once per year.

\begin{figure}
\centering
\includegraphics[width=0.4\textwidth, keepaspectratio, angle=270]{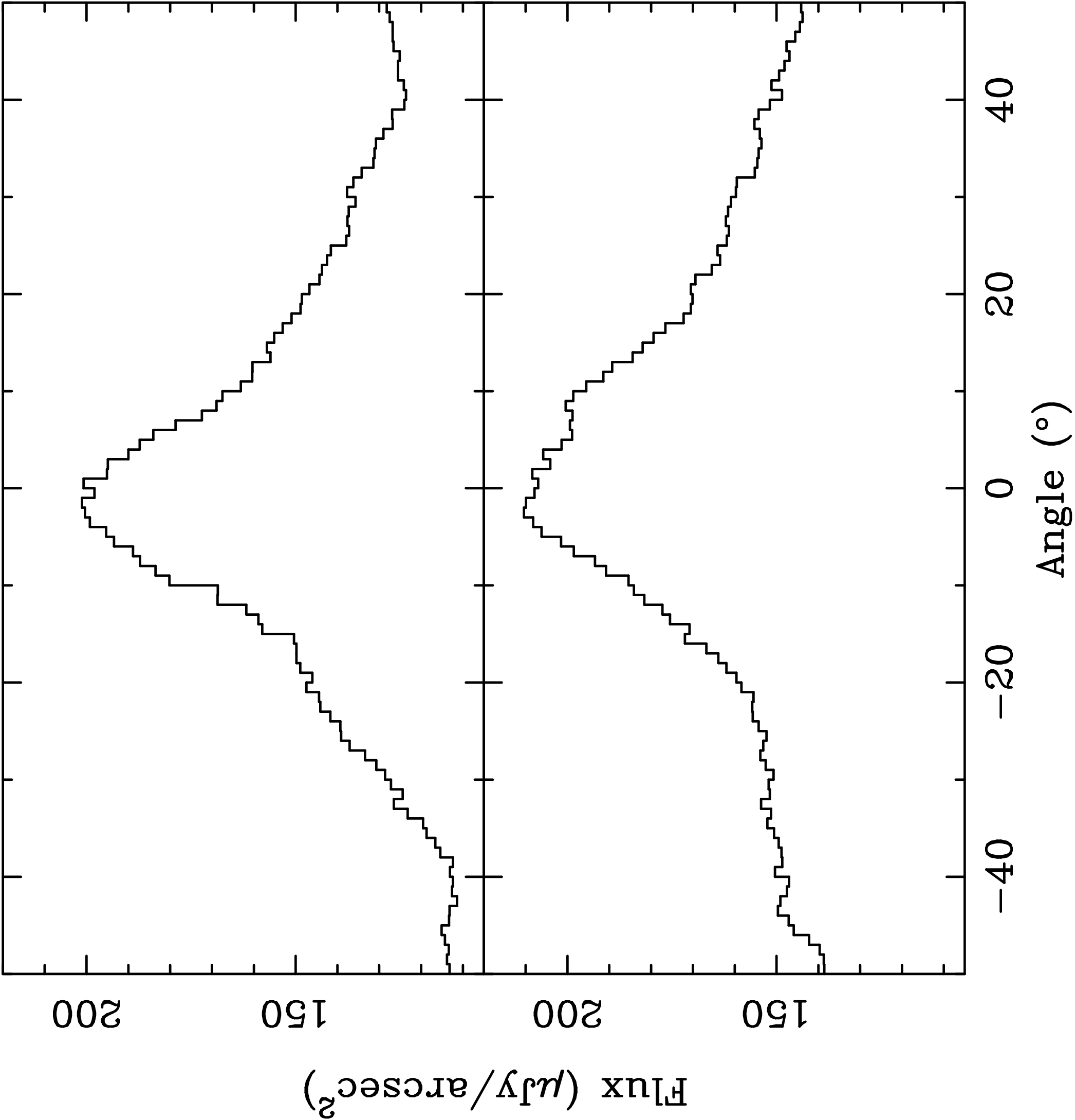}
\caption{The azimuthal distributions of two optical wisps observed on
  2011  April 13 at distances of $8.0\:  arcsec$ and
  $10.1\:  arcsec$ from the pulsar. The angle is the azimuthal angle of
  the deprojected ellipse.
$0^{\circ}$ was set by the peak of the distribution.}
\label{fig:Crab_angular_projection}
\end{figure}

\subsection{Analysis of the azimuthal profile of the wisps} \label{s:wap}

Good statistics for the optical data allow us to measure the azimuthal intensity profile for each observation. These analyses measure the azimuthal intensity distribution of the flux in the ellipse-like shapes to the NW of the pulsar. 
Elliptical shapes arise presumably because they are formed in a circular ring more or less in the equatorial plane of the pulsar.
In X-rays one finds that the aspect ratio of the innermost ellipse
(the location of the presumed termination shock) is 0.49 and the
pulsar is displaced along the axis of symmetry by $\approx 0.9\:  arcsec$ below
the plane of the ring and along the minor axis which is at a
position angle of $-60\degree$ \citep{Weisskopf2012}.
It is more difficult to determine these parameters in the visible
primarily because a complete ring is not observed.
Nevertheless, as may be seen in Fig.~\ref{fig:Crab_NOT}, assuming an elliptical shape with the offset and aspect ratio as specified follows the path of the optical data extremely well.
Since we have already found that the wisps in both bands typically do
not line up, there is no
reason to expect that the ellipses be identical.

Fig.~\ref{fig:Crab_angular_projection} shows the azimuthal
distribution of the intensity for two of the optical wisps at
distances of $8.0\:  arcsec$and $10.1\:  arcsec$ from the pulsar.
The distributions shown in the figure were based on
 a $0.5\:  arcsec$-wide elliptical annulus with $0.6$ aspect ratio.

The X-ray data are noisier than the optical data so we
considered both the average of the 2011 April data and the
average of all except for the 2011 April data.
Within errors, the width of these two distributions were identical.
The azimuthal variation was obtained from an
elliptical annulus $0.5\:  arcsec$ wide and $8.0\:  arcsec$ from the pulsar as shown in Fig.~\ref{fig:Crab_Chandra}.

Comparing the X-ray angular distribution with the optical one, it is
important to note that the latter has a sharper peak than that seen in X-rays.
The full width at half-maximum (FWHM) of the distribution in optical is around $30^\circ$ while the
 X-ray distribution has an FWHM of around $70^\circ$.

\begin{figure*}
{\includegraphics[width=0.45\textwidth]{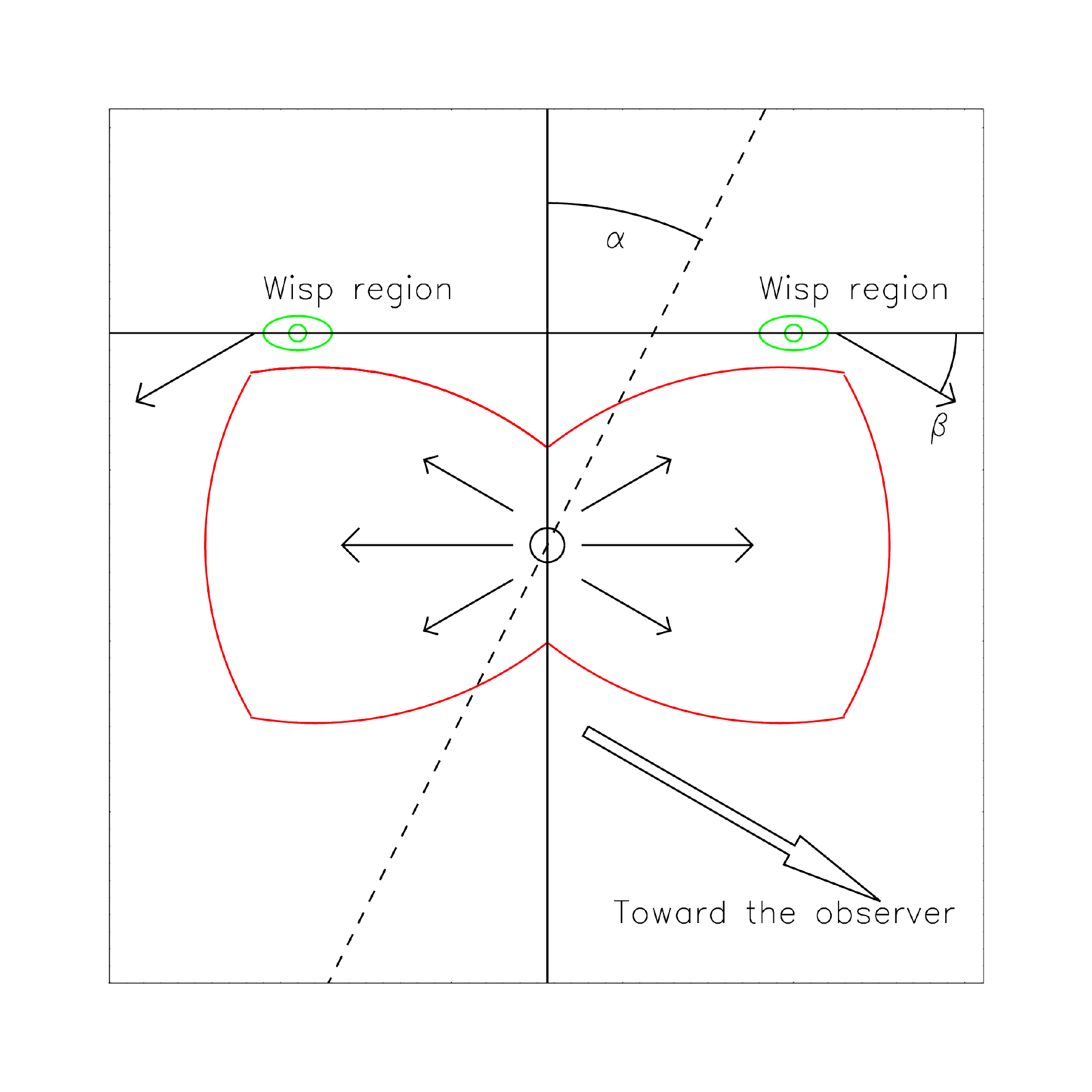}\hspace{0cm} \includegraphics[width=0.45\textwidth]{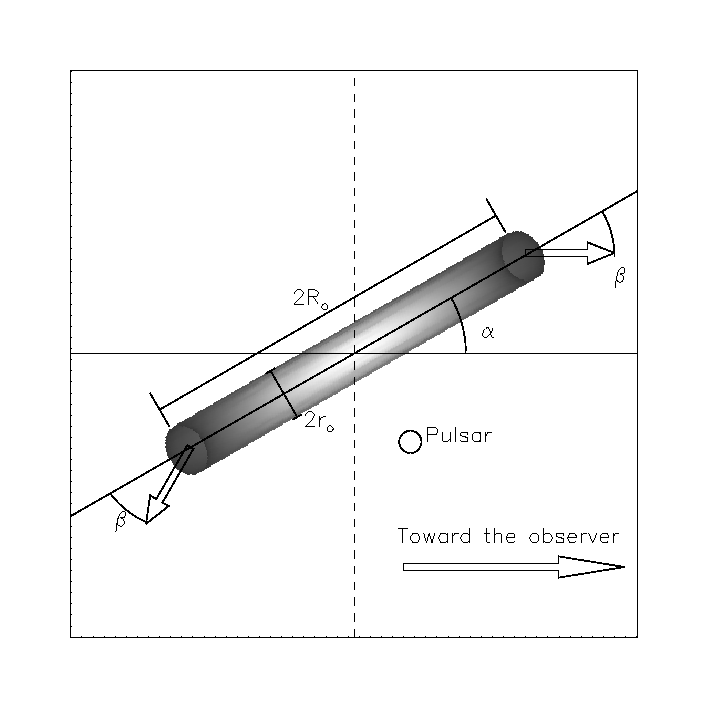}}
\caption{
Left-hand panel: Schematic structure of the termination shock
(red) cross-section.
The pulsar, from which a relativistic radial wind originates, is
located at the central circle
and the solid vertical line represents the axis of symmetry of the nebula.
The large arrow indicates the direction towards the observer, while
the dashed line represents the plane of the sky.
The wind slows down and is diverted in a termination shock which has an
oblique shape due to wind anisotropy. 
Wisps can form anywhere along the termination shock.
A wisp (solid green circle) will appear bright within the wisp region (solid green ellipse) when the particle motion is directed
towards the observer. The wisp  will fade as it moves out from this region.
The symbol $\alpha$ is the angle between the nebular axis and the
plane of the sky and $\beta$ is the angle between the plane of the
wisps and the flow direction in the wisp regions. Right-hand panel: Schematic of a single wisp modelled as
an emitting ring. The dashed line represents the plane of the sky and $R_{o}$ and
$r_{o}$ are the major and minor radii of the torus respectively. 
}
\label{fig:model}
\end{figure*}

\section{Model fitting and theoretical discussion} \label{s:mft}

The magnetohydrodynamic (MHD) model developed with success in the past years
\citep{Komissarov_Lyubarsky03a,Komissarov_Lyubarsky04a,Del-Zanna_Amato+04a,Bucciantini_del-Zanna+05a,Del-Zanna_Volpi+06a,Volpi_Del-Zanna+08b,Camus_Komissarov+09a} explains the existence of rings and wisps observed in pulsar wind nebulae (PWNe) in terms of axisymmetric enhancements of the emissivity immediately downstream of the termination shock.
Due to the anisotropy of the pulsar wind luminosity, and the related
oblate shape of the termination shock \citep{Del-Zanna_Amato+04a}, axisymmetric regions form inside the nebula where the emissivity towards the observer is higher, either because of local compression of the magnetic field, or because of Doppler boosting.
In general it is found that typical flow speeds immediately downstream of the termination shock can be as high as $0.7$c, and the emission is dominated by Doppler boosting.

\citet{Komissarov_Lyubarsky04a} have shown that both the wisps and the knot observed in the Crab nebula can be qualitatively explained by the geometry of the flow (Fig.~\ref{fig:model}).
In particular the association of the wisps with the wind termination shock is due to the fact that they originate in the flow immediately downstream of the
shock itself.
A simplified model for a wisp can be built on the assumption that a wisp is due to a torus- or ring-like region within the nebula.
This region can appear bright or dim depending on Doppler boosting due to the relative direction of the observer, and the particle flow.
This boosted-ring model appears in alternative explanations for the origin of the wisps, e.g. in term of ion-cyclotron compression \citep{Spitkovsky_Arons04a} or cooling instabilities \citep{Foy_Hester09a}.
The boosting is required in all of these models to explain the azimuthal luminosity profile of the wisps, and the fact that one side of the nebula (the front side) is brighter than the back side.
This idea has been adopted in the past \citep{Romani_Ng03a,Ng_Romani04a,Romani_Ng+05a} to model the jet torus structure observed in X-rays in several PWNe, and to estimate typical flow speeds in these systems.

\begin{table}
\begin{center}
\begin{tabular}{lcccccr}
\hline \hline
  MJD       & Dist & $V_{fl}$ &  $U_{fl}$ & $V_{fl,min}$ & $V_{fl,max}$ &
  $r_o/R_o$\vspace{0.1cm}\\
\hline
55518 & 5.0      & 0.87	& 1.76          & 0.80          & 0.92  &       0.01\\
55518 & 7.2	 & 0.96	& 3.68          & 0.96          & 0.97  &       0.01\\
55518 & 8.0      & 0.95	& 2.89          & 0.93          & 0.96  &       0.01\\
55518 & 9.7	 & 0.80	& 1.33          & 0.75          & 0.85  &       0.01\\
55532 & 7.7	 & 0.94	& 2.76          & 0.93          & 0.95  &       0.01\\
55532 & 10.1     & 0.83 & 1.49          & 0.75	        & 0.87  &	0.01\\
55557 & 8.0	 & 0.94	& 2.76          & 0.93		& 0.95	&       0.01\\
55557 & 10.9     & 0.90 & 2.06          & 0.84		& 0.93	&	0.04\\
55564 & 8.1	 & 0.95	& 3.04          & 0.93		& 0.96	&	0.01\\
55564 & 11.2     & 0.92	& 2.35          & 0.88		& 0.96	&	0.04\\
55570 & 8.2 	 & 0.95	& 3.04          & 0.94		& 0.96	&	0.01\\
55570 & 11.4     & 0.92	& 2.35          & 0.87		& 0.94	&	0.03\\
55598 & 8.9	 & 0.94	& 2.76          & 0.92		& 0.95	&	0.04\\
55598 & 11.1     & 0.90	& 2.06          & 0.85		& 0.92	&	0.03\\
55598 & 12.7     & 0.70	& 0.98          & 0.60		& 0.85	&	0.04\\
55612 & 8.4	 & 0.93	& 2.53          & 0.90		& 0.95	&	0.03\\
55612 & 9.1 	 & 0.93	& 2.53          & 0.90		& 0.94	&	0.01\\
55612 & 11.4     & 0.80	& 1.33          & 0.70		& 0.87	&	0.02\\
55647 & 7.6	 & 0.97	& 3.99          & 0.95		& 0.98	&	0.01\\
55647 & 9.7 	 & 0.97	& 3.99          & 0.96		& 0.98	&	0.01\\
55647 & 12.4     & 0.90	& 2.06          & 0.93		& 0.85	&	0.01\\
55664 & 5.8	 & 0.85	& 1.61          & 0.77		& 0.91	&	0.04\\
55664 & 8.0	 & 0.93	& 2.53          & 0.91		& 0.95	&	0.04\\
55664 & 10.1     & 0.91	& 2.19          & 0.87		& 0.93	&	0.04\\
55810 & 6.9	 & 0.93	& 2.53          & 0.90	        & 0.95	&	0.07\\
55810 & 12.5     & 0.93	& 2.53          & 0.89	        & 0.96	&	0.06\\
55863 & 5.0      & 0.80	& 1.33          & 0.73		& 0.90	&	0.02\\
55863 & 7.6	 & 0.93	& 2.53          & 0.90		& 0.95	&	0.07\\
55863 & 12.5     & 0.93	& 2.53          & 0.90		& 0.95	&	0.06\\
55881 & 7.2 	 & 0.95	& 3.04          & 0.94		& 0.96	&	0.01\\
55881 & 8.0	 & 0.94	& 2.75          & 0.93		& 0.95	&	0.01\\
55881 & 10.2     & 0.80	& 1.33          & 0.70		& 0.87	&	0.04\\
55881 & 12.6     & 0.90	& 2.06          & 0.85		& 0.93	&	0.06\\
55922 & 6.0	 & 0.86	& 1.68          & 0.82		& 0.90	&	0.02\\
55922 & 8.2	 & 0.93	& 2.53          & 0.90		& 0.97	&	0.04\\
55954 & 8.6 	 & 0.90	& 2.06          & 0.89		& 0.93	&	0.07\\
55954 & 12.5     & 0.90	& 2.06          & 0.85		& 0.95	&	0.07\\
55998 & 9.6	 & 0.92	& 2.35          & 0.89		& 0.95	&	0.10\\
56013 & 7.6	 & 0.93	& 2.53          & 0.90		& 0.95	&	0.03\\
56013 & 8.9	 & 0.92	& 2.35          & 0.89		& 0.94	&	0.03\\
56025 & 7.5	 & 0.94	& 2.75          & 0.92		& 0.97  &	0.04\\
56025 & 9.3	 & 0.90	& 2.06          & 0.86		& 0.92	&	0.03\\
56033 & 7.6	 & 0.94	& 2.75          & 0.92		& 0.97	&	0.04\\
56033 & 9.6	 & 0.87	& 1.76          & 0.85		& 0.89	&	0.04\\
56140 & 8.0	 & 0.95	& 3.04          & 0.93		& 0.96	&	0.05\\
56140 & 13.6     & 0.85	& 1.61          & 0.80		& 0.88	&	0.10\\
56141 & 8.0	 & 0.95	& 3.04          & 0.93		& 0.96	&	0.05\\
56141 & 13.8     & 0.80	& 1.33          & 0.70		& 0.85	&	0.10\\
56153 & 8.3	 & 0.96	& 3.43          & 0.94		& 0.97	&	0.02\\
56159 & 8.3	 & 0.95	& 3.04          & 0.93		& 0.96	&	0.06\\
56176 & 8.6	 & 0.95	& 3.04          & 0.93		& 0.96	&	0.06\\

\end{tabular}
\caption{Fitted parameters for the optical wisps. The table lists the MJD, the distance to the pulsar, the flow velocity in units of c, the boosting four-velocity, and the minimum and maximum flow velocities still compatible with the fit and the thickness of the torus. In all cases we found $\alpha=\beta= 37^\circ\pm4^\circ$. }
\end{center}
\label{tab:fit}
\end{table}

With reference to Fig.~\ref{fig:model}, a wisp originates from a region shaped like a torus, with major radius $R_o$ and a minor radius $r_o$.
The ratio $r_o/R_o$ we call the thickness of the torus.
The plane of the torus has an inclination $\alpha$ with respect to the
line of sight, which, within the axisymmetric approximation, is equal to the angle between the nebular axis and the plane of the sky.
The fluid in the torus has a uniform flow speed $V_{fl}$, confined in
meridianal planes (planes containing the nebular axis).
The azimuthal component of this flow speed is assumed to be $0$.
$V_{fl}$ forms an angle $\beta$ with respect to the plane of the torus.

We want to stress here that $V_{fl}$ is the flow speed of particles in the wisp, and should not be confused with the observed wisp velocity, which we have discussed previously.
The two are not the same simply because the wisp velocity traces the propagation of the ring in the nebula as a wave, and not the bulk motion of particles.
For simplicity we assume uniform synchrotron emissivity inside the torus.
In this picture the torus contains a fully ordered toroidal magnetic field, but models can also be built with a chaotic magnetic field (see below for a discussion).
The emissivity towards the observer is computed as in previous works \citep{Komissarov_Lyubarsky03a,Komissarov_Lyubarsky04a, Del-Zanna_Amato+04a,Bucciantini_del-Zanna+05a,Del-Zanna_Volpi+06a,Volpi_Del-Zanna+08b,Camus_Komissarov+09a} and a synchrotron map is built integrating the
contribution of the various parts of the torus along the line of sight.
The emitting particles are assumed to be distributed as a power-law in energy with an index $-2.35$ in agreement with values estimated from optical observations and spectral modelling of the Crab nebula \citep{Bucciantini_Arons+11a}.
This also applies for the X-ray wisp(s).
In the MHD model of PWNe, the wisps are assumed to be located immediately downstream of the termination shock, where particles are accelerated and magnetic field is compressed.
Given that the wisps originate from the termination shock, synchrotron cooling should be negligible and the particle distribution unaffected.

The wisps appear brighter to the NW of the pulsar and almost fade to
the sky background level in the southeast mostly due to Doppler
boosting, given that the flow velocity is generally directed towards
the observer in the NW, while in the southeast it points away.
We want to stress here that the plane of the torus does not necessarily need to coincide (even if it is parallel) with the equatorial plane of the pulsar, but there can be an offset between the two.
Such an offset, however, has no consequences in the model for the wisp emissivity.

\begin{figure*}
\centering
{\includegraphics[width=0.45\textwidth]{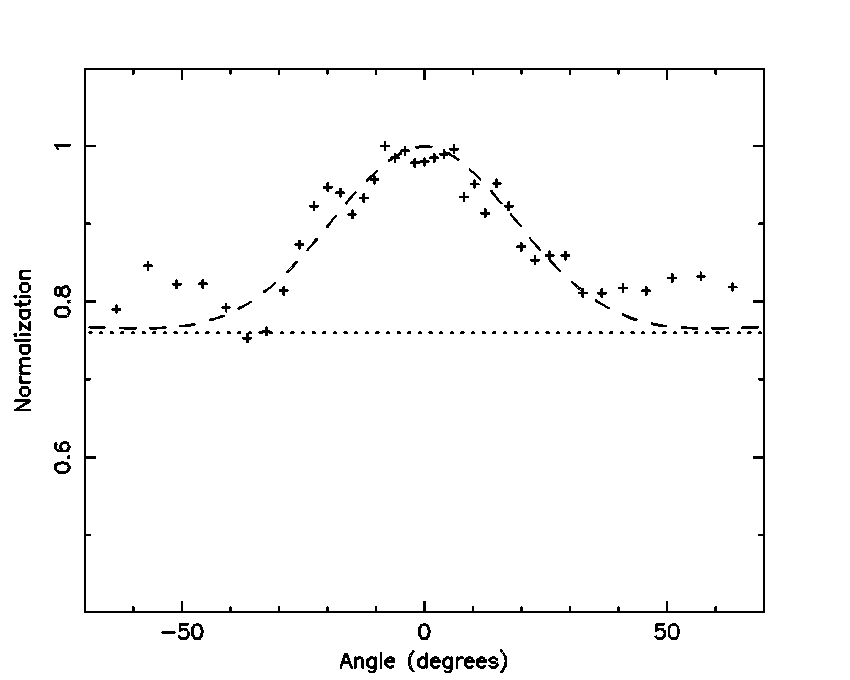} \includegraphics[width=0.45\textwidth]{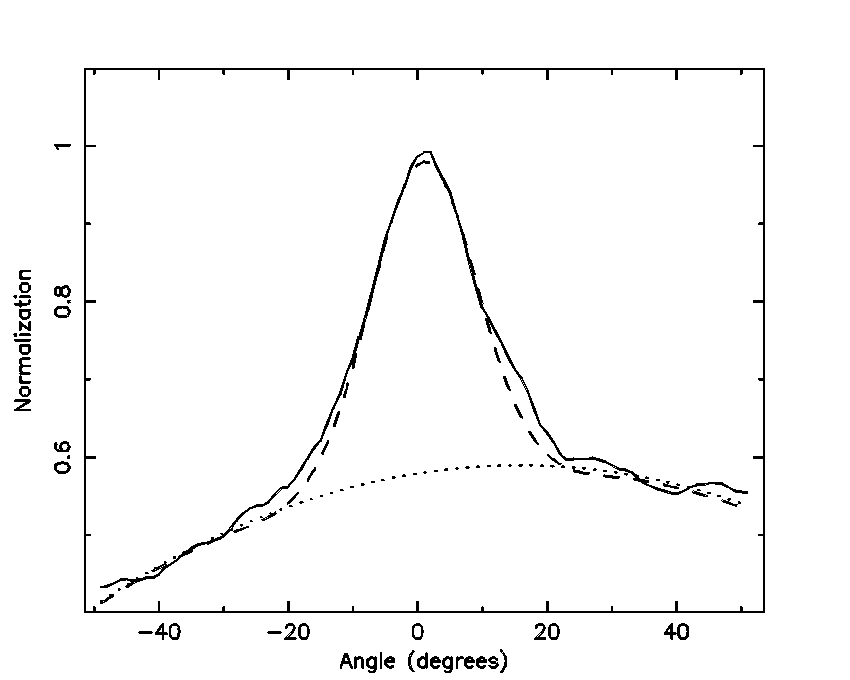}}
\caption{Left-hand panel: The average azimuthal X-ray intensity
  distribution (normalized to its maximum) in an elliptical annulus at
  a distance of $8\:  arcsec$ to the NW of the pulsar.
The dashed line shows a simulated azimuthal distribution as discussed in \S\ref{s:mft}.
The dotted line denotes a``background'' level based on a simple 2D 
polynomial  model for the local azimuthal intensity.
The dashed line shows a simulated azimuthal distribution as discussed in \S\ref{s:mft}.
 Right-hand panel: Same as the left-hand panel but for the average optical
wisp centred at a distance of $8 arcsec$ from the pulsar.
}
\label{fig:xofit}
\end{figure*}

\begin{figure}
{\includegraphics[width=0.4\textwidth]{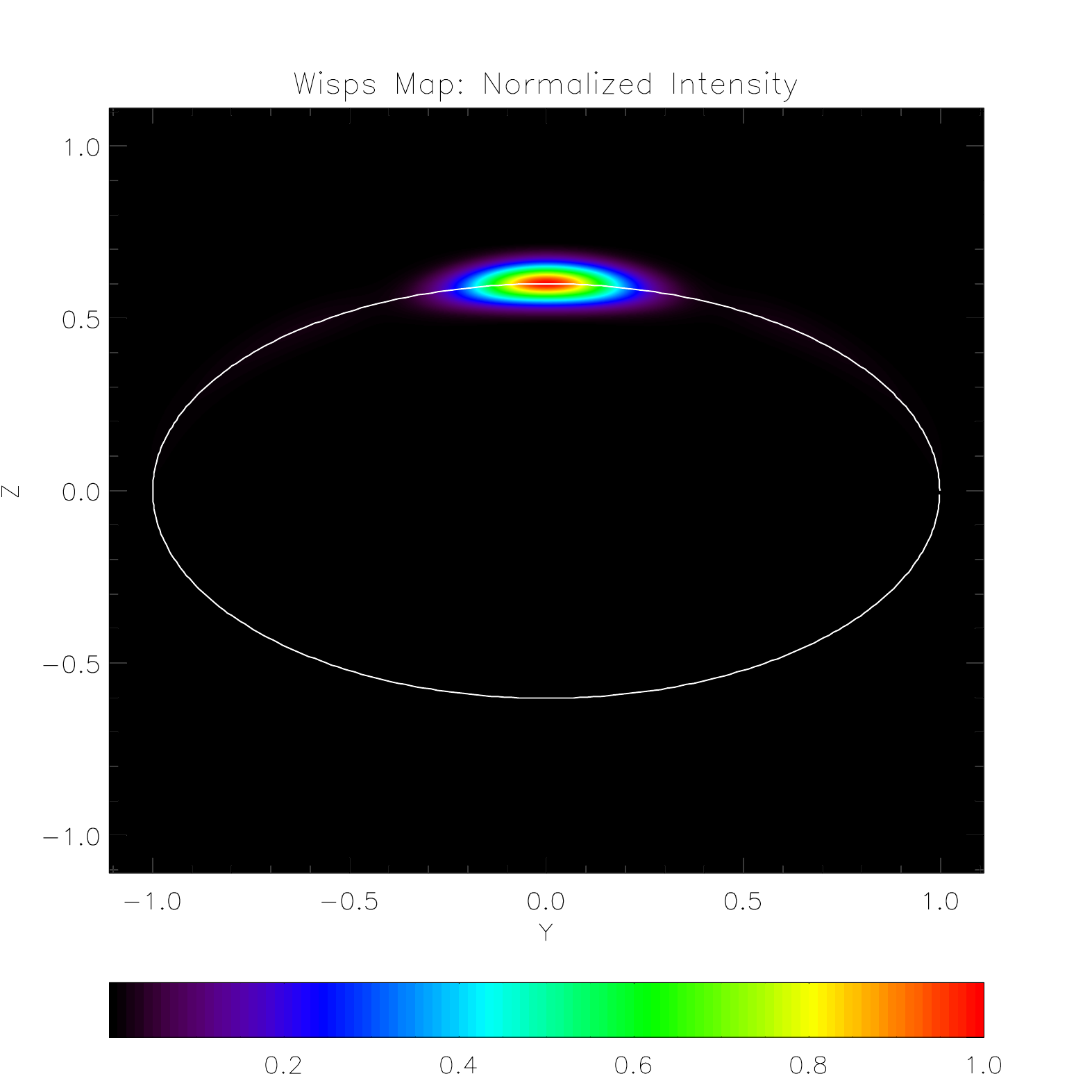} \includegraphics[width=0.4\textwidth]{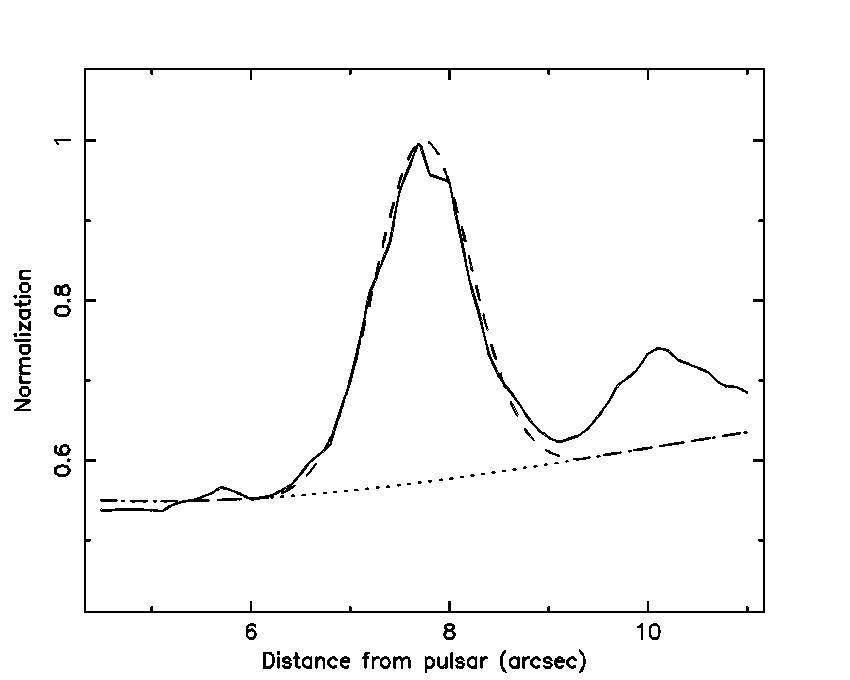} \includegraphics[width=0.4\textwidth]{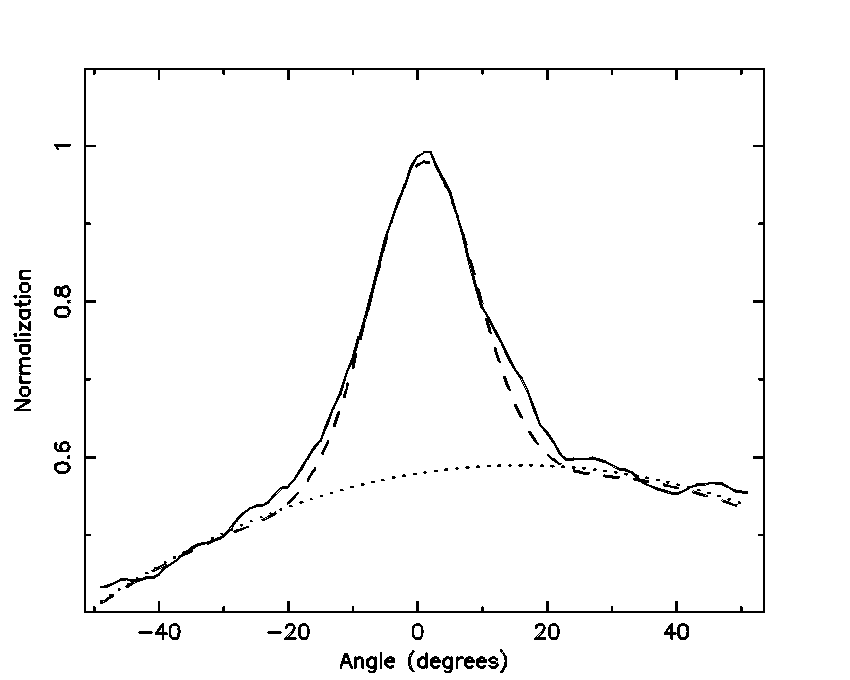}}
\caption{Upper panel: Simulated synchrotron map of the optical wisp for MJD=55532 and at a distance of $7.7\:  arcsec$ derived using the fitted parameters in Table~\ref{tab:fit}.
The axes are in arbitrary units normalized to the wisp major radius $R_o$. Colours indicate the level of the flux, normalized to the maximum.
Middle panel: the solid line is the optical profile of the wisp along the nebular axis, while the dashed line is obtained from the simulated synchrotron map.
The dotted line represents the (assumed) background.
Lower panel: Same as the middle panel but for the azimuthal profile.
}
\label{fig:example_fit}
\end{figure}

We fit the optical data to this model as follows:
\begin{itemize}
\item We select a set of input parameters $r_o/R_o$, $\alpha$, $\beta$, and
$V_{fl}$.
\item An emission map is built (the emission is normalized to the maximum).
\item The emission map is convolved with the PSF of the observation.
\item Emission profiles are extracted both along the wisp and along
the axis of the nebula.
\item

The simulated radial and azimuthal profiles are added to an assumed
background and then compared with the data.
The background is based on a simple 2D polynomial least-squares fit,
as shown in Fig.~\ref{fig:xofit} and Fig.~\ref{fig:example_fit}.
Note that we found no significant difference in the results if
fits were performed to the background subtracted data.
Furthermore, there were no significant differences in the boosting speed if the background in the wisp region is assumed to be uniform, or to have low frequency components.

\item This procedure is repeated until a set of input parameters are found that provide a reasonable fit.
The best-fitting parameters are obtained by minimizing the residuals in the brightest part of the wisp, where the background subtracted intensity is $>50$\% of its maximum value.
\end{itemize}

The results for the optical wisps are presented in Table~\ref{tab:fit}.
We can also roughly quantify the uncertainty in the various parameters.
The upper (lower) limit on the flow boosting velocity $V_{fl}$ given in Table~\ref{tab:fit}, corresponds to a model which gives a wisp whose FWHM is 15\%
narrower (larger) than the observed value.
Fig.~~\ref{fig:example_fit} compares one of the optical observations and the result from a simulated map, using the best-fitting input parameters from Table~\ref{tab:fit}.

We stress here that our procedure does not take into account time-of-flight
delays.
However, given that the extent of the optical wisps is usually within $\pm 20^\circ$ from the axis, this effect should be negligible.

There are several interesting points to note:
\begin{itemize}
\item Typical flow speeds inferred from our model range between $V_{fl}=0.8$c and $V_{fl}=0.95$c.
Using a chaotic magnetic field distribution leads to even higher velocities.
\item The angle $\beta$ is almost equal to the angle $\alpha$.
\item The wisps appear to be unresolved and the ratio $r_o/R_o$ is less than $0.1$.
\end{itemize}

\begin{figure}
{\includegraphics[width=0.5\textwidth]{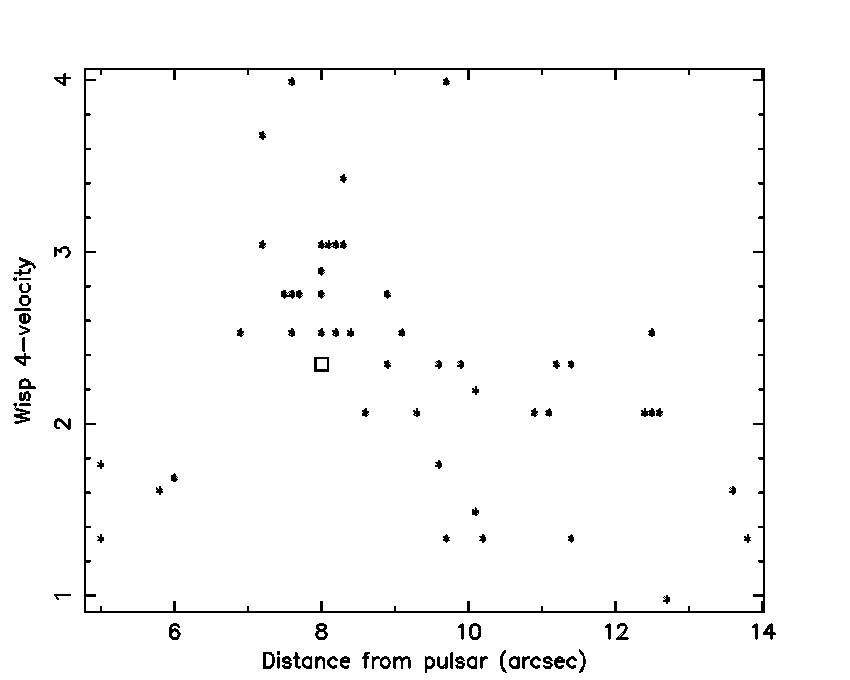}}
\caption{Boosting four-velocity $(V_{fl}/c)/\sqrt{[1-(V_{fl}/c)^2]}$
  for the optical wisps as a function of their distance from the pulsar.
The square denotes the boosting four-velocity obtained from the averaged optical
profile. Note that the data points at radii $\leq7\:  arcsec$ are most
likely  not due to wisps, but other optical features such as the halo
\citep{Hester08}, in which case the model would not apply. 
}
\label{fig:4velocity}
\end{figure}

The high values of the flow speed correspond to boosting factors that
are in excess of typical values found in MHD simulations by about a
factor of between 2 and 5 depending on which value of $\beta$ one chooses.
This is  related to the narrow extent of the optical wisps which fade to the sky background level within $\pm 20^\circ$ from the axis.
This is also connected to the fact that $\beta \approx\alpha$, implying that the flow speed on the front side of the nebula is directed towards the observer.
If $\beta$ differs from $\alpha$ by more than $3^\circ-5^\circ$, it is
not possible to reproduce the azimuthal wisp luminosity, independently
of $V_{fl}$. The minimum value of $V_{fl}$ is obtained for $\beta=\alpha$.
Moreover, it appears that the optical wisps are consistent with narrow features, possibly close to emitting sheets.
It is also interesting to note that there is a trend in the observed flow
speeds as a function of wisp location, with higher values for inner
wisps, as shown in Fig.~\ref{fig:4velocity}.

A comparison between the azimuthal profile in optical and X-rays has also
been done.
However, for the X-ray data, in order to have sufficient
statistics, instead of a single epoch, we consider first an average over
the entire set of observations and only for the inner (brighter) X-ray wisp
region located $\sim 8\:  arcsec$ away from the pulsar (Fig.~~\ref{fig:xofit}).
We also repeated this analysis averaging just the 2011 April data with similar results.

To eliminate the possibility that the difference was due to averaging
the X-ray data, we averaged the optical data over all of our epochs.
Results were shown in Fig.~\ref{fig:xofit}.
The resulting model parameters are $r_o/R_o=0.1$,
$\alpha=\beta=35^\circ$ for both optical and X-ray, while
$V_{fl}=0.6c\pm 0.1c$ for the X-ray and $V_{fl}=0.91c\pm 0.03c$ for
the optical.
We may associate this with a boosting four-velocity $U_{fl}=(V_{fl}/c)/\sqrt{[1-(V_{fl}/c)^2]}$
of 0.75 for X-ray and 2.20 for optical (i.e. a factor of 3.13).
Thus, we can safely conclude that the azimuthal extent of this X-ray
wisp is larger than in the optical
and, within the context of our model, this implies lower inferred speeds for the particles producing the X-rays.
The fitted parameters for the optical wisps, however, raise the question if the boosted ring model is in fact appropriate.
The MHD model has been developed mostly to reproduce Chandra observations but 
it is also evident from our study that X-ray and optical wisps are not produced by the same particle distribution: they do not coincide in location or in terms of the degree of Doppler boosting.
It is not our intention here to develop and provide an alternative explanation for the azimuthal luminosity of the optical wisps, but we can point to a few possible solutions.
\begin{itemize}
\item Optical wisps appear to originate in the transition layer at the termination shock itself.
This is consistent with the thin-sheet hypothesis and the idea that, at the shock itself, the flow speed is higher than in the downstream region.
It remains to be proven that the geometry of the shock can allow for such configurations.
\item Emissivity is enhanced along the axis of the nebula because of local clumpiness.
Clumps are observed in X-rays, called by some {\it sprites}, and they have a more or less stable location.
However, why in optical, these clumps should be located along the axis of the nebula is not clear.
\item If the particle distribution in a wisp is non-isotropic, then
  the   emissivity has a stronger dependence on the inclination of the magnetic field with respect to the line of sight than in the isotropic case.
It is not clear, however, what the degree of anisotropy needs to be and if it can be produced by known acceleration mechanisms.
\end{itemize}

\section{Conclusion}\label{s:sr}

We find that approximately once per year optical and X-ray wisps appear and peel off from the region commonly associated with the termination shock of the pulsar wind.
Although the X-ray and optical wisps are related by their proximity and their outward motion, they are not precisely aligned.
It is also interesting to note that the optical and radio wisps are separated from each other \citep{Bietenholz04}.

Interestingly, the time interval between $\gamma$-ray flares is also about 1 y leading to the speculation that the flaring may somehow be associated with wisp formation.
However, since a new wisp
did not form in coincidence with the $\gamma$-ray flare of 2011 April which took place during this study, the speculation requires a theory that has the formation of the wisps either substantially ($\approx$month) preceding a $\gamma$-ray flare, or being an unrelated phenomenon.

The wisps propagate outward with velocities projected on to the sky ranging from $ 0.1$(v/c) to $0.4$(v/c) (Table~\ref{tab:velocities}).
If anything, there appears to be a trend for the projected velocity of a wisp increasing, the further it is from the pulsar, perhaps implying a re-acceleration pmechanism for the wisp itself.
More likely, this apparent behaviour may be a result of the more complicated three-dimensional geometry and there may be no need for such re-acceleration when properly deprojected.
A word of caution is that one might be confusing features in the counter-jet with structure in the wisps.

Within the context of an MHD modelling, we find that optical wisps are
more strongly Doppler-boosted than the X-ray wisps.
In particular, we found that the azimuthal luminosity profile of the
X-ray wisps is fully compatible with typical boosting factors found
in MHD simulations of PWNe.
Instead, the azimuthal luminosity profile of the optical wisps
requires particles velocities that are incompatible with the results
of global numerical modelling of PWNe
\citep{Komissarov_Lyubarsky03a,Bucciantini_del-Zanna+05a}.
This should be investigated with future modeling.

\section*{Acknowledgments}
The optical observations used in this work have been done with the Nordic Optical
Telescope, operated on the island of La Palma jointly by Denmark,
Finland, Iceland, Norway and Sweden, in the Spanish Observatorio del
Roque de los Muchachos of the Instituto de Astrofisica de Canarias.
The data were obtained [in part] with ALFOSC, which is provided by the
Instituto de Astrofisica de Andalucia (IAA) under a joint agreement
with the University of Copenhagen and NOTSA.
The staff at the NOT provided much useful support, particularly
Thomas Augusteijn.

The X-ray observations have been done with the Chandra X-ray satellite.
A.T. and M.C.W would like to acknowledge support from the Chandra
Project. We would also like to acknowledge the Director of the Chandra
Science Center for authorizing Director's Discretionary Time and
Chandra Proposal 1350025 for the remainder of the observations.

We thank the referee for many insightful and clarifying comments.

\bibliography{my}{}
\bibliographystyle{mn2e}

\label{lastpage}
\end{document}